\documentclass[12pt]{article}
\usepackage{graphicx}
\usepackage{amsmath}
\usepackage{amssymb}
\usepackage{caption2}
\begin{document}
\bibliographystyle{prsty}
\begin{center}
{\large {\bf \sc{  Is   $D_s(2700)$ a charmed tetraquark state?  }}} \\[2mm]
Zhi-Gang Wang \footnote{E-mail,wangzgyiti@yahoo.com.cn.  }     \\
 Department of Physics, North China Electric Power University,
Baoding 071003, P. R. China

\end{center}

\begin{abstract}
In this article, we take the point of view that the  $D_s(2700)$ be
a  tetraquark state, which consists of  a scalar diquark and a
vector antidiquark, and calculate its mass with the QCD sum rules.
The numerical result
  indicates that the mass of the vector charmed tetraquark state is
  about $M_V=(3.75\pm0.18)\rm{GeV}$ or $M_V=(3.71\pm0.15)\rm{GeV}$ from different sum rules,
  which is about
  $1\rm{GeV}$ larger than the experimental data. Such  tetraquark
  component should be very small in the $D_s(2700)$.

\end{abstract}



\section{Introduction}

Recently   Belle Collaboration observed  a new resonance
$D_{s}(2700)$
 in the decay $B^{+} \to \bar{D}^{0} D_{s}(2700) \to \bar{D}^{0}
D^{0} K^{+}$. The resonance has  the mass  $M_V=2708 \pm 9
^{+11}_{-10} \rm{MeV}$, width $\Gamma_V = 108 \pm 23 ^{+36}_{-31}
~\rm{MeV}$, and spin-parity $1^{-}$\cite{Belle}. They interpret the
 $D_s (2700)$ as a $c\bar{s}$ meson, the potential model
calculations predict a  radially excited $2^3S_1$ ($c\bar{s}$) state
with a mass about $(2710-2720)\rm{MeV}$ \cite{Pmodel}. The resonance
$D_{s}(2700)$ is consistent with the particle they presented
previously in the 33rd international conference on high energy
physics (ICHEP 06), $M_V = 2715 \pm 11^{+11}_{-14} \rm{MeV}$, $
\Gamma_V= 115 \pm 20^{+36}_{-32} \rm{MeV}$ and spin-parity $1^-$
\cite{Belle2}.   In the same analysis of the $DK$ mass distribution,
Babar Collaboration observed a broad structure with $M_V = 2688 \pm
4 \pm 3 \rm{MeV}$ and $\Gamma_V = 112\pm 7\pm 36 \rm{MeV}$, which
maybe the same resonance observed by Belle Collaboration
\cite{Babar} .

In this article, we take the point of view that the vector charmed
meson $D_{s}(2700)$ be a tetraquark state, which  consists of a
scalar diquark and a vector antidiquark, and devote to calculate its
mass with the QCD sum rules \cite{SVZ79,Narison89}. The $D_s(2700)$
lies above the $DK$ threshold,  the decay $ D_{s}(2700) \to D^{0}
K^{+}$ can take place with the fall-apart mechanism and it is OZI
super-allowed, which can take into account the large width
naturally. Furthermore, whether or not there exists such a
tetraquark configuration  which can result in the state
$D_{s}(2700)$ is of great importance itself, because it provides a
new opportunity for a deeper understanding of low energy QCD. We
explore this possibility, later experimental data can confirm or
reject this assumption.

The article is arranged as follows:  we derive the QCD sum rules for
the mass  of  the $D_s(2700)$  in section 2; in section 3, numerical
result and discussion; section 4 is reserved for conclusion.

\section{QCD sum rules for the mass of the $D_s(2700)$ }
In the following, we write down  the two-point correlation function
$\Pi_{\mu\nu}(p)$ in the QCD sum rules,
\begin{eqnarray}
\Pi_{\mu\nu}(p)&=&i\int d^4x e^{ip \cdot x} \langle
0|T\left\{J_\mu(x)J^+_\nu(0)\right\}|0\rangle \, ,  \\
J_\mu(x)&=&\frac{\epsilon_{kij}\epsilon_{kmn}}{\sqrt{2}}\left\{u^T_i(x)C\gamma_5
c_j(x) \bar{u}_m(x)\gamma_5 \gamma_\mu C
\bar{s}^T_n(x)+(u\rightarrow d) \right\}\, .
\end{eqnarray}
We  choose the vector current $J_\mu(x)$ which  constructed from a
scalar diquark and a vector antidiquark  to interpolate the vector
meson $D_s(2700)$.

Here we take a digression to discuss  how to choose the
interpolating currents for the tetraquark  states.  We can take
either $qq-\bar{q}\bar{q}$ type or
 $\bar{q}q-\bar{q}q$ type  currents to interpolate  the tetraquark
 states, they are related to each other via Fierz transformation both in the Dirac
 spinor and color space \cite{Zhu06,Buballa05}.
In this article, we take the  $qq-\bar{q}\bar{q}$ type interpolating
current.

The diquarks have  five Dirac tensor structures, scalar $C\gamma_5$,
pseudoscalar $C$, vector $C\gamma_\mu \gamma_5$, axial vector
$C\gamma_\mu $  and  tensor $C\sigma_{\mu\nu}$. From those diquarks,
we can construct six independent currents to interpolating the
charmed tetraquark states with $1^-$,
\begin{eqnarray}
J^1_\mu(x)&=&\frac{\epsilon_{kij}\epsilon_{kmn}}{\sqrt{2}}\left\{u^T_i(x)C\gamma_5
c_j(x) \bar{u}_m(x)\gamma_5 \gamma_\mu C
\bar{s}^T_n(x)+(u\rightarrow d) \right\}\, , \nonumber\\
J^2_\mu(x)&=&\frac{\epsilon_{kij}\epsilon_{kmn}}{\sqrt{2}}\left\{u^T_i(x)C
\gamma_\mu \gamma_5c_j(x) \bar{u}_m(x)\gamma_5 C
\bar{s}^T_n(x)+(u\rightarrow d) \right\}\, , \nonumber\\
J^3_\mu(x)&=&\frac{\epsilon_{kij}\epsilon_{kmn}}{\sqrt{2}}\left\{u^T_i(x)C
c_j(x) \bar{u}_m(x)  \gamma_\mu C \bar{s}^T_n(x)+(u\rightarrow d)
\right\}\, , \nonumber\\
J^4_\mu(x)&=&\frac{\epsilon_{kij}\epsilon_{kmn}}{\sqrt{2}}\left\{u^T_i(x)C
\gamma_\mu c_j(x) \bar{u}_m(x)    C \bar{s}^T_n(x)+(u\rightarrow d)
\right\}\, , \nonumber\\
J^5_\mu(x)&=&\frac{\epsilon_{kij}\epsilon_{kmn}}{\sqrt{2}}\left\{u^T_i(x)C\sigma_{\mu\nu}
c_j(x) \bar{u}_m(x)\gamma_5 \gamma_\nu C
\bar{s}^T_n(x)+(u\rightarrow d) \right\}\,, \nonumber \\
J^6_\mu(x)&=&\frac{\epsilon_{kij}\epsilon_{kmn}}{\sqrt{2}}\left\{u^T_i(x)C\gamma_\nu
\gamma_5 c_j(x) \bar{u}_m(x)\sigma_{\mu\nu} C
\bar{s}^T_n(x)+(u\rightarrow d) \right\}\,  ,
\end{eqnarray}
and the general current $\widetilde{J}_\mu(x)$ can be written as
their linear superposition,
\begin{eqnarray}
\widetilde{J}_\mu(x)&=& \sum_{i=1}^6C_iJ^i_\mu(x) \, ,
\end{eqnarray}
where the $C_i$ are some coefficients.

 The six interpolating currents can be sorted into three types, the
currents $J^1_\mu(x)$ and $J^2_\mu(x)$ are $C\gamma_5-C\gamma_\mu
\gamma_5$ type, the currents $J^3_\mu(x)$ and $J^4_\mu(x)$ are
$C-C\gamma_\mu $ type, the currents $J^5_\mu(x)$ and $J^6_\mu(x)$
are $C\sigma_{\mu\nu}-C\gamma_\nu \gamma_5$ type. We expect the
three  type interpolating  currents result in three type  of masses
for the tetraquark states.

The study with the  random instanton liquid model indicates that the
diquarks have masses about $m_S=420\pm30\rm{MeV}$,
$m_A=m_V=940\pm20\rm{MeV}$, $m_T=570\pm20\rm{MeV}$ \cite{Schafer94},
we expect the currents $J^5_\mu(x)$ and $J^6_\mu(x)$  interpolate
the tetraquark states with masses larger than the ones for the
currents $J^1_\mu(x)$ and $J^2_\mu(x)$. Instanton induced force
results in strong attraction in the scalar diquark channels and
strong repulsion in the pseudoscalar diquark channels, if the
instantons manifest themselves, the pseudoscalar diquarks will have
much larger masses than the corresponding scalar diquarks
\cite{GluonInstanton}, the coupled Schwinger-Dyson equation and
Bethe-Salpeter equation also indicate this fact \cite{Burden97}.
Furthermore, the one-gluon exchange force leads to significant
attraction between the quarks in the $0^+$ channels
\cite{GluonInstanton}. Although the interpolating currents are not
unique, the currents $J^1_\mu(x)$ and $J^2_\mu(x)$ are much better
and interpolate tetraquark states with smaller mass, we can choose
either one of them.

In the conventional QCD sum rules \cite{SVZ79}, there are two
criteria (pole dominance and convergence of the operator product
expansion) for choosing  the Borel parameter $M^2$ and threshold
parameter $s_0$. For the tetraquark states, if the perturbative
terms have the main contribution, we can approximate the spectral
density with the perturbative term,
\begin{eqnarray}
B_M\Pi \sim A \int_0^\infty s^4 e^{-\frac{s}{M^2}}ds=A
M^{10}\int_0^\infty t^4 e^{-t}dt \, ,
\end{eqnarray}
where the $A$ are some numerical coefficients, then we take the pole
dominance condition,
\begin{eqnarray}
\frac{\int_0^{t_0} t^4 e^{-t}dt}{\int_0^\infty t^4 e^{-t}dt}\geq
50\% \, ,
\end{eqnarray}
and obtain the relation,
\begin{eqnarray}
t_0&=&\frac{s_0}{M^2}\geq 4.7 \, .
\end{eqnarray}
The superpositions of different interpolating currents can only
change the contributions from different terms in the operator
product expansion, and  improve convergence, they cannot change the
leading behavior of the spectral density $\rho(s)\propto s^4 $ of
the perturbative term.

For the nonet light  scalar mesons below $1\rm{GeV}$, if their
dominant Fock components are tetraquark states,  even we choose
special superposition of different currents to weaken the
contributions from the vacuum condensates to warrant the main
contribution from  the perturbative term, we cannot choose very
small Borel parameter $M^2$ to enhance  the pole term. For small
enough Borel parameter $M^2$, the perturbative corrections of order
$\mathcal {O}(\alpha_s(M))$, $\mathcal {O}(\alpha_s^2(M))$,
$\cdots$, maybe large enough to invalidate the operator product
expansion.

We can choose the typical energy scale $\mu=M=1\rm{GeV}$, in that
energy scale, $\sqrt{s_0}\approx 2.2\rm{GeV}$. There are many scalar
mesons below $2\rm{GeV}$ \cite{PDG}, their contributions are already
included in at the phenomenological side. Pole dominance cannot be
fully satisfied for the  tetraquark states with light flavor.

Failure of pole dominance do not mean non-existence of the
tetraquark states, it just means that the QCD sum rules, as one of
the QCD models, may have shortcomings.  We release  some  criteria
and take more phenomenological analysis, i.e. we choose larger Borel
parameter $M^2$ to warrant   convergence of the operator product
expansion  and take a phenomenological cut off  to avoid  possible
comminations from the high resonances and continuum states
\cite{WangTetraquark}.

If we insist on to retain pole dominance besides convergence of the
operator product expansion in the QCD sum rules for the tetraquark
states, the hidden charmed and bottomed tetraquark states,  and open
bottomed tetraquark states may satisfy the criterion in Eq.(7), as
they always have larger Borel parameter $M^2$ and threshold
parameter $s_0$.

For examples, in Ref.\cite{Narison07}, the authors take the
$X(3872)$ as hidden charmed tetraquark state and calculate its mass
with the QCD sum rules, the Borel parameter and threshold parameter
are taken as $M^2=(2.0-2.8)\rm{GeV^2}$ and $s_0=(17-18)\rm{GeV}^2$;
in Ref.\cite{Nielsen07}, the authors take the $Z(4430)$ as hidden
charmed tetraquark state and calculate its mass with
$M^2=(2.5-3.1)\rm{GeV^2}$ and $s_0=(23-25)\rm{GeV}^2$, furthermore,
the authors  calculate the corresponding bottomed one, and choose
$M^2=(8.0-8.3)\rm{GeV^2}$ (or $M^2=(8.0-9.9)\rm{GeV^2}$) and
$s_0=125\rm{GeV}^2$(or $s_0=135\rm{GeV}^2$). In those sum rules,
although the windows for the Borel parameters are rather small, the
$\alpha_s(M)$ is small enough to warrant  convergence of the
operator product expansion, the relation in Eq.(7) can be well
satisfied.

The correlation function  $\Pi_{\mu\nu}(p)$ can be decomposed as

\begin{eqnarray}
\Pi_{\mu\nu}(p)&=&- \Pi_1(p^2)\left\{g_{\mu\nu}-\frac{p_\mu
p_\nu}{p^2} \right\} +\Pi_0(p^2)\frac{p_\mu p_\nu}{p^2},
\end{eqnarray}
due to  Lorentz covariance.  The invariant functions $\Pi_1$ and
$\Pi_0$ stand for the contributions from the vector and scalar
mesons, respectively. In this article, we choose the tensor
structure $g_{\mu\nu}-\frac{p_\mu p_\nu}{p^2}$ to study the mass of
the vector meson.

 According to   basic assumption of current-hadron duality in
the QCD sum rules \cite{SVZ79}, we insert  a complete series of
intermediate states satisfying  unitarity   principle with the same
quantum numbers as the current operator $J_\mu(x)$
 into the correlation function $\Pi_{\mu\nu}(p)$  to obtain the hadronic representation. After isolating the
pole term  of the lowest state $D_s(2700)$, we obtain the following
result:
\begin{eqnarray}
\Pi_{\mu\nu}(p)&=&-\frac{f_{V}^2M_{V}^8}{M_{V}^2-p^2}\left\{g_{\mu\nu}-\frac{p_\mu
p_\nu}{p^2} \right\} +\cdots \, \, ,
\end{eqnarray}
where we have used the following definition,
\begin{eqnarray}
\langle 0|J_\mu(0) |D_s(2700)\rangle&=&f_{V}M_{V}^4 \epsilon_\mu \,
,
\end{eqnarray}
here $\epsilon_\mu$ is the  polarization vector of the $D_s(2700)$
and $f_V$ is the residue of the pole.

In the following, we briefly outline  operator product expansion for
the correlation function $\Pi_{\mu\nu }(p)$  in perturbative QCD
theory. The calculations are performed at   large space-like
momentum region $p^2\ll 0$, which corresponds to small distance
$x\approx 0$ required by   validity of   operator product expansion.
We write down the "full" propagators $S_{ij}(x)$(the $U_{ij}(x)$ and
$D_{ij}(x)$ for the $u$ and $d$ quarks can be obtained with a simple
replacement of the nonperturbative parameters) and $C_{ij}(x)$ of a
massive quark in the presence of the vacuum condensates firstly
\cite{SVZ79}\footnote{One can consult the last article of
Ref.\cite{SVZ79} for technical details in deriving the full
propagator.},
\begin{eqnarray}
S_{ij}(x)&=& \frac{i\delta_{ij}\!\not\!{x}}{ 2\pi^2x^4}
-\frac{\delta_{ij}m_s}{4\pi^2x^2}-\frac{\delta_{ij}}{12}\langle
\bar{s}s\rangle +\frac{i\delta_{ij}}{48}m_s
\langle\bar{s}s\rangle\!\not\!{x}-\nonumber\\
&&\frac{\delta_{ij}x^2}{192}\langle \bar{s}g_s\sigma Gs\rangle
 +\frac{i\delta_{ij}x^2}{1152}m_s\langle \bar{s}g_s\sigma
 Gs\rangle \!\not\!{x}-\nonumber\\
&&\frac{i}{32\pi^2x^2} G^{ij}_{\mu\nu} (\!\not\!{x}
\sigma^{\mu\nu}+\sigma^{\mu\nu} \!\not\!{x})  +\cdots \, ,
\end{eqnarray}
\begin{eqnarray}
C_{ij}(x)&=&\frac{i}{(2\pi)^4}\int d^4k e^{-ik \cdot x} \left\{
\frac{\delta_{ij}}{\!\not\!{k}-m_c}
-\frac{g_sG^{\alpha\beta}_{ij}}{4}\frac{\sigma_{\alpha\beta}(\!\not\!{k}+m_c)+(\!\not\!{k}+m_c)\sigma_{\alpha\beta}}{(k^2-m_c^2)^2}\right.\nonumber\\
&&\left.+\frac{\pi^2}{3} \langle \frac{\alpha_sGG}{\pi}\rangle
\delta_{ij}m_c \frac{k^2+m_c\!\not\!{k}}{(k^2-m_c^2)^4}
+\cdots\right\} \, ,
\end{eqnarray}
where $\langle \bar{s}g_s\sigma Gs\rangle=\langle
\bar{s}g_s\sigma_{\alpha\beta} G^{\alpha\beta}s\rangle$  and
$\langle \frac{\alpha_sGG}{\pi}\rangle=\langle
\frac{\alpha_sG_{\alpha\beta}G^{\alpha\beta}}{\pi}\rangle$, then
contract the quark fields in the correlation function
$\Pi_{\mu\nu}(p)$ with Wick theorem, and obtain the result:
\begin{eqnarray}
\Pi_{\mu\nu}(p)&=&i \epsilon_{kij}\epsilon_{k'i'j'}
\epsilon_{kmn}\epsilon_{k'm'n'}\int d^4x \, e^{i p \cdot x}
Tr\left\{ \gamma_5\gamma_\mu CS_{n'n}^T(-x)C\gamma_\nu \gamma_5
U_{m'm}(-x)\right\} \nonumber \\
&&Tr\left\{\gamma_5 C_{jj'}(x)\gamma_5 C U^T_{ii'}(x)C\right\}\, .
\end{eqnarray}
Substitute the full $s$, $c$ and $u$ quark propagators into above
correlation function and complete  the integral in  coordinate
space, then integrate over the variable $k$, we can obtain the
correlation function $\Pi_1(p^2)$ at the level of quark-gluon
degrees  of freedom:

\begin{eqnarray}
\Pi_1(p^2)&=& -\frac{1}{61440\pi^6} \int_0^1 dt
\left[\mathcal{K}^4(\frac{7}{t^3}+\frac{3}{t^2})+4\mathcal{K}^3 p^2
(1+\frac{3}{t}-\frac{4}{t^2}) \right]\log\mathcal{K}
\nonumber \\
&&-\frac{m_s p^2}{192\pi^4}\int_0^1 dt \left[6(t-1)\langle
\bar{q}q\rangle+(t^2+t-2)\langle
\bar{s}s\rangle\right] \mathcal{K}\log \mathcal{K} \nonumber \\
&&+\frac{m_c \langle \bar{q}q\rangle }{192\pi^4}\int_0^1 dt
 (\frac{1}{t}+\frac{2}{t^2})\mathcal{K}^2\log \mathcal{K} \nonumber \\
&&-\frac{m_s }{128\pi^4}\int_0^1 dt \left[\frac{4}{t}\langle
\bar{q}q\rangle+(1+\frac{1}{t})\langle
\bar{s}s\rangle\right] \mathcal{K}^2\log \mathcal{K} \nonumber \\
&&+\frac{m_c\langle \bar{q}g_s\sigma Gq\rangle}{128\pi^4} \int_0^1
dt(1+\frac{1}{t} )\mathcal{K} \log \mathcal{K}\nonumber \\
&&-\frac{m_s}{384\pi^4} \int_0^1 dt \left[(3t+1)\langle
\bar{s}g_s\sigma Gs\rangle+12\langle
\bar{q}g_s\sigma Gq\rangle \right]\mathcal{K} \log \mathcal{K}\nonumber \\
&&+\frac{m_sp^2}{384\pi^4} \int_0^1 dt \left[(t-t^3)\langle
\bar{s}g_s\sigma Gs\rangle+6(t-t^2)\langle
\bar{q}g_s\sigma Gq\rangle \right] \log \mathcal{K}\nonumber \\
&&-\frac{\langle \bar{q}q\rangle \langle \bar{s}s\rangle}{6\pi^2}
\int_0^1 dt\mathcal{K}\log \mathcal{K}
+\frac{\langle \bar{q}q\rangle \langle \bar{s}s\rangle p^2}{12\pi^2} \int_0^1 dt (t-t^2)\log \mathcal{K}\nonumber \\
&&+\frac{m_c m_s}{24\pi^2} \int_0^1 dt \left[ 2\langle \bar{q}q\rangle^2+t \langle \bar{q}q\rangle \langle \bar{s}s\rangle\right]\log \mathcal{K}\nonumber \\
&&-\frac{\langle \bar{q}q\rangle \langle \bar{s}g_s\sigma Gs\rangle+
\langle \bar{s}s\rangle\langle \bar{q}g_s\sigma Gq\rangle }{24\pi^2}
\int_0^1 dt t\log \mathcal{K} \,\, ,
\end{eqnarray}
where $\mathcal{K}(p^2)=(1-t)m_c^2-t(1-t)p^2$.

  We carry out  operator
product expansion to the vacuum condensates adding up to
dimension-8. In calculation, we
 take  assumption of vacuum saturation for  high
dimension vacuum condensates, they  are always
 factorized to lower condensates with vacuum saturation in the QCD sum rules,
  factorization works well in  large $N_c$ limit.
In this article, we take into account the contributions from the
quark condensates,  mixed condensates, and neglect the contributions
from the gluon condensate. In calculation, we observe the
contributions  from the gluon condensate are suppressed by large
denominators and would not play any significant roles.

Once  analytical results are obtained,
  then we can take  current-hadron duality  below the threshold
$s_0$ and perform  Borel transformation with respect to the variable
$P^2=-p^2$, finally we obtain  the following sum rules:

\begin{eqnarray}
f_{V}^2M_{V}^8\exp\left\{-\frac{M_V^2}{M^2}\right\}&=&\int_{m_c^2}^{s_0}
ds
\frac{\mbox{Im}\Pi(s)}{\pi} \exp\left\{-\frac{s}{M^2}\right\} \, ,\\
M_{V}^2&=&\int_{m_c^2}^{s_0} ds \frac{\mbox{Im}\Pi(s)}{\pi}s
\exp\left\{-\frac{s}{M^2}\right\} /\nonumber\\
&&\int_{m_c^2}^{s_0} ds \frac{\mbox{Im}\Pi(s)}{\pi}
\exp\left\{-\frac{s}{M^2}\right\} \, ,
\end{eqnarray}
\begin{eqnarray}
\frac{\mbox{Im}\Pi(s)}{\pi}&=& \frac{1}{61440\pi^6} \int_\Delta^1 dt
\left[\mathcal{K}^4(\frac{7}{t^3}+\frac{3}{t^2})+4\mathcal{K}^3 s
(1+\frac{3}{t}-\frac{4}{t^2}) \right]
\nonumber \\
&&+\frac{m_s s}{192\pi^4}\int_\Delta^1 dt \left[6(t-1)\langle
\bar{q}q\rangle+(t^2+t-2)\langle
\bar{s}s\rangle\right] \mathcal{K}  \nonumber \\
&&-\frac{m_c \langle \bar{q}q\rangle }{192\pi^4}\int_\Delta^1 dt
 (\frac{1}{t}+\frac{2}{t^2})\mathcal{K}^2  \nonumber\\
 &&+\frac{m_s }{128\pi^4}\int_\Delta^1 dt \left[\frac{4}{t}\langle
\bar{q}q\rangle+(1+\frac{1}{t})\langle
\bar{s}s\rangle\right] \mathcal{K}^2 \nonumber \\
&&-\frac{m_c\langle \bar{q}g_s\sigma Gq\rangle}{128\pi^4}
\int_\Delta^1
dt(1+\frac{1}{t} )\mathcal{K} \nonumber \\
&&+\frac{m_s}{384\pi^4} \int_\Delta^1 dt \left[(3t+1)\langle
\bar{s}g_s\sigma Gs\rangle+12\langle
\bar{q}g_s\sigma Gq\rangle \right]\mathcal{K}  \nonumber \\
&&-\frac{m_s s}{384\pi^4} \int_\Delta^1 dt \left[(t-t^3)\langle
\bar{s}g_s\sigma Gs\rangle+6(t-t^2)\langle
\bar{q}g_s\sigma Gq\rangle \right]  \nonumber  \\
&&+\frac{\langle \bar{q}q\rangle \langle \bar{s}s\rangle}{6\pi^2}
\int_\Delta^1 dt\mathcal{K}
-\frac{\langle \bar{q}q\rangle \langle \bar{s}s\rangle s}{12\pi^2} \int_\Delta^1 dt (t-t^2) \nonumber \\
&&-\frac{m_c m_s}{24\pi^2} \int_\Delta^1 dt \left[ 2\langle \bar{q}q\rangle^2+t \langle \bar{q}q\rangle \langle \bar{s}s\rangle\right] \nonumber \\
&&+\frac{\langle \bar{q}q\rangle \langle \bar{s}g_s\sigma Gs\rangle+
\langle \bar{s}s\rangle\langle \bar{q}g_s\sigma Gq\rangle }{24\pi^2}
\int_\Delta^1 dt t \, ,
\end{eqnarray}
where  $\Delta=\frac{m_c^2}{s}$.

\section{Numerical result and discussion}
The input parameters are taken to be the standard values $\langle
\bar{q}q \rangle=-(0.24\pm 0.01 \rm{GeV})^3$, $\langle \bar{s}s
\rangle=(0.8\pm 0.2 )\langle \bar{q}q \rangle$, $\langle
\bar{q}g_s\sigma Gq \rangle=m_0^2\langle \bar{q}q \rangle$, $\langle
\bar{s}g_s\sigma Gs \rangle=m_0^2\langle \bar{s}s \rangle$,
$m_0^2=(0.8 \pm 0.2)\rm{GeV}^2$,    $m_s=(0.14\pm0.01)\rm{GeV}$ and
$m_c=(1.4\pm0.1)\rm{GeV}$ \cite{SVZ79,Narison89,Ioffe2005}. For the
multiquark  states,  the contribution from  terms with the gluon
condensate $\langle \frac{\alpha_s GG}{\pi}\rangle $ is of minor
importance \cite{WangTetraquark}, and the contribution from the
$\langle \frac{\alpha_s GG}{\pi}\rangle $ is neglected here.

\begin{table}
\begin{center}
\begin{tabular}{c|c}
\hline\hline
      $\mbox{perturbative term}$  &$+96\%$\\ \hline
      $\langle \bar{q} q\rangle$, $\langle \bar{s} s\rangle$& $+33\%$\\      \hline
     $\langle \bar{q}g_s\sigma G q\rangle$, $\langle \bar{s} g_s
\sigma G s\rangle$& $-10\%$ \\     \hline
    $\langle \bar{q} q\rangle^2$,$\langle \bar{q} q\rangle\langle
\bar{s} s\rangle$&  $-24\%$\\ \hline $\langle \bar{q} q\rangle
\langle \bar{s}g_s \sigma G s\rangle$, $\langle \bar{q}g_s \sigma G
q\rangle\langle \bar{s} s\rangle$& $+4\%$\\ \hline \hline
\end{tabular}
\end{center}
\caption{ The contributions from different terms in Eq.(15) for
$s_0=16\rm{GeV}^2$ and $M^2=6\rm{GeV}^2$. }
\end{table}

From  Table 1, we can see that the dominating contribution comes
from the perturbative term, (a piece of)  standard criterion of the
QCD sum rules can be satisfied. If we change the Borel parameter in
the interval $M^2=(5-7)\rm{GeV}^2$, the contributions from different
terms change slightly.

Although the contributions from the terms concerning  the quark
condensates and mixed condensates are rather large, however, they
are canceled out with each other, the net contributions are of minor
importance. Which is in contrast to the sum rules with other
interpolating currents constructed from the multiquark
configurations, where the contribution comes from the perturbative
term is very small \cite{Narison04}, the main contributions come
from the terms with the quark condensates $\langle \bar{q}q \rangle$
and $\langle \bar{s}s\rangle$, sometimes the mixed condensates
$\langle \bar{q}g_s\sigma G q \rangle$ and $\langle \bar{s}g_s\sigma
G s\rangle$ also play important roles (for example, the first three
articles of the Ref.\cite{WangTetraquark}). One can choose special
superposition of different currents to weaken the contributions from
the vacuum condensates to warrant the main contribution from the
perturbative term, it is somewhat of fine-tuning.

The values of the vacuum condensates have been updated with the
experimental data for  $\tau$ decays, the QCD sum rules for the
baryon masses and analysis of the charmonium spectrum
\cite{Ioffe2005}. As the main contribution comes from the
perturbative term, uncertainties of the vacuum condensates can only
result in very small uncertainty for numerical value of the mass
$M_V$, the standard values and updated values of the vacuum
condensates can only lead to results of minor difference, we choose
the standard values of the vacuum condensates in the calculation.

In Fig.1, we plot the value of the  $M_V$ with  variations of the
threshold parameter $s_0$ and Borel parameter $M^2$.  If
$\sqrt{s_0}\leq 3.55\rm{GeV}$, $M_V > s_0$, we  cannot take into
account all   contributions  from the $D_s(2700)$, furthermore, the
$M_V$ changes quickly with the variation of the Borel parameter
$M^2$, the threshold parameter $s_0$ should be chosen to be
$\sqrt{s_0}>3.6\rm{GeV}$. The value of the $M_V$ is almost
independent on the Borel parameter $M^2$ at about
$\sqrt{s_0}=4.0\rm{GeV}$. In this article, the threshold parameter
$s_0$ is chosen to be $s_0=(16\pm2)\rm{GeV}^2$. It is large enough
for the Breit-Wigner mass $M_V=2708 \pm 9 ^{+11}_{-10} \rm{MeV}$,
width $\Gamma_V = 108 \pm 23 ^{+36}_{-31} ~\rm{MeV}$. However, the
standard criterion of pole dominance cannot be satisfied, the
contribution from the pole term is less than $13\%$. If one insist
on that the multiquark states should satisfy the same criteria as
the conventional mesons  and baryons, the QCD sum rules for the
(light and charmed) tetraquark states should be discarded. For
detailed discussions about how to select the Borel parameters and
threshold parameters for the multiquark states, one can consult
Ref.\cite{WangTetraquark}.

\begin{figure}
 \centering
 \includegraphics[totalheight=7cm,width=13cm]{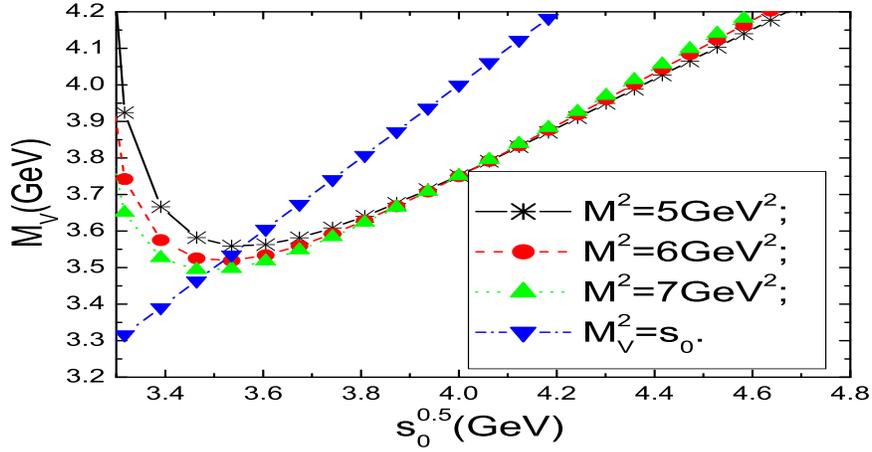}
  \caption{ $M_V$ with  Borel parameter $M^2$ and threshold parameter $s_0$. }
\end{figure}

\begin{figure}
 \centering
 \includegraphics[totalheight=7cm,width=13cm]{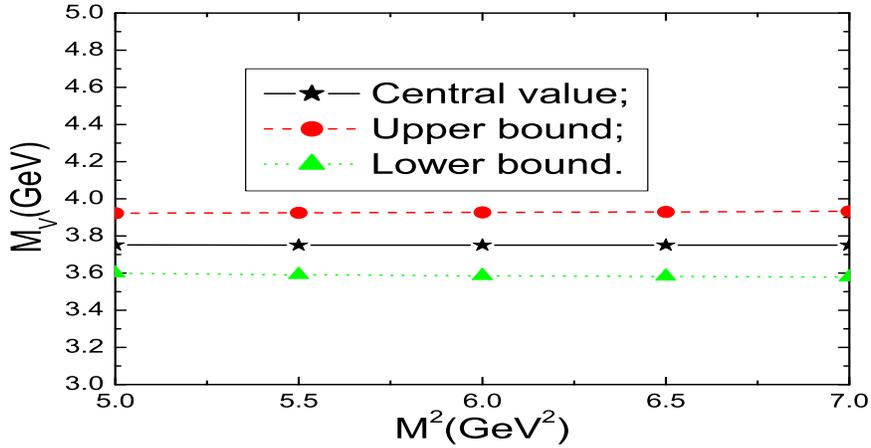}
  \caption{ $M_V$ with  Borel parameter $M^2$ from Eq.(16). }
\end{figure}

Taking into account all the uncertainties, we obtain the value of
the mass of the $D_s(2700)$, which is shown in Fig.2,
\begin{eqnarray}
M_V&=&(3.75\pm0.18)\rm{GeV} \,   .
\end{eqnarray}

It is obvious that our numerical value is larger than the
experimental data $M_V=2.708\rm{GeV}$, the vector current can
interpolate a charmed tetraquark state with the mass about
$M_V=3.75\rm{GeV}$ or even larger, such tetraquark component should
be small in the $D_s(2700)$.

If one want to retain the pole dominance of the conventional QCD sum
rules, we take the replacement for the weight functions in
Eqs.(15-16),
\begin{eqnarray}
\exp\left\{-\frac{s}{M^2}\right\} &\rightarrow&
\exp\left\{-\left(\frac{s}{M^2}\right)^2\right\} \, , \nonumber\\
\exp\left\{-\frac{M_V^2}{M^2}\right\} &\rightarrow&
\exp\left\{-\left(\frac{M_V^2}{M^2}\right)^2\right\} \, ,
\end{eqnarray}
and obtain new QCD sum rules for the mass of the vector tetraquark
state.
\begin{eqnarray}
f_{V}^2M_{V}^8\exp\left\{-\left(\frac{M_V^2}{M^2}\right)^2\right\}&=&\int_{m_c^2}^{s_0}
ds
\frac{\mbox{Im}\Pi(s)}{\pi} \exp\left\{-\left(\frac{s}{M^2}\right)^2\right\} \, ,\\
M_{V}^4&=&\int_{m_c^2}^{s_0} ds \frac{\mbox{Im}\Pi(s)}{\pi}s^2
\exp\left\{-\left(\frac{s}{M^2}\right)^2\right\} /\nonumber\\
&&\int_{m_c^2}^{s_0} ds \frac{\mbox{Im}\Pi(s)}{\pi}
\exp\left\{-\left(\frac{s}{M^2}\right)^2\right\} \, ,
\end{eqnarray}

As the main contributions come from the perturbative term,
  the hadronic spectral density above and below the threshold
can be successfully approximated by the perturbative term. If we
take typical values for the parameters $\sqrt{s_0}=4.0\rm{GeV}$ and
$M^2=(7-9)\rm{GeV}^2$, the contribution from  pole term in Eq.(20)
is dominating, about $53\%-84\%$. Taking into account all the
uncertainties, we obtain the value of the mass of $D_s(2700)$, which
is shown in Fig.3,
\begin{eqnarray}
M_V&=&(3.71\pm0.15)\rm{GeV} \,   .
\end{eqnarray}

\begin{figure}
 \centering
 \includegraphics[totalheight=7cm,width=13cm]{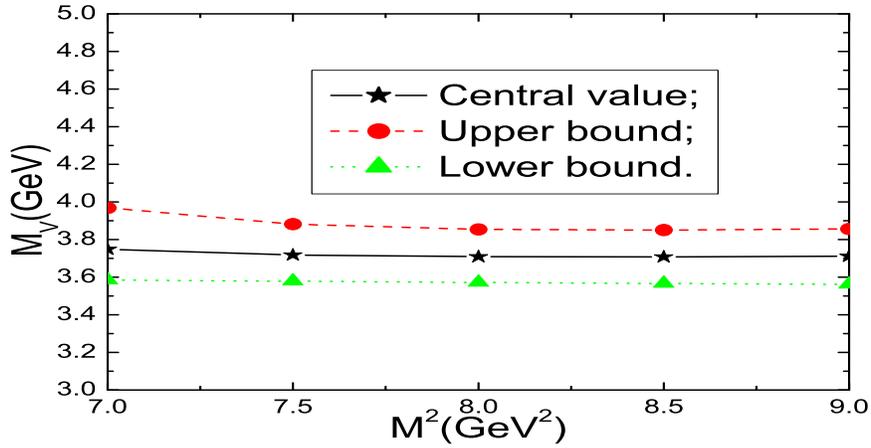}
  \caption{ $M_V$ with  Borel parameter $M^2$ from Eq.(21). }
\end{figure}
\section{Conclusion}
In this article, we take the point of view that  the $D_s(2700)$ be
a tetraquark state which consists of  a scalar diquark and a vector
antidiquark, and calculate its mass with
  the QCD sum rules.  The numerical result
  indicates that the mass of vector charmed tetraquark state is about
  $M_V=(3.75\pm0.18)\rm{GeV}$ or $M_V=(3.71\pm0.15)\rm{GeV}$, which is about
  $1\rm{GeV}$ larger than the experimental data. Such
  tetraquark component  should be
  very small in the $D_s(2700)$, the dominating component may be the $c\bar{s}$ state, we
  can take up the method developed in Ref.\cite{Oka07} to study the mixing between the
  two-quark component and tetraquark component with the interpolating
  current $\widehat{J}_\mu(x)=cos\theta J_\mu(x)+sin \theta\langle \bar{q}q\rangle \bar{s}(x)\gamma_\mu c(x)$.
  The decay  $D_{s}(2700) \to
D^{0} K^{+}$ can occur mainly through  creation of the $u\bar{u}$
pair in the QCD vacuum, we resort to the $^3P_0$ model to calculate
the decay width \cite{3P0}, although the $^3P_0$ model is rather
crude.

\section*{Acknowledgments}
This  work is supported by National Natural Science Foundation,
Grant Number 10405009, 10775051, and Program for New Century
Excellent Talents in University, Grant Number NCET-07-0282.

\end{document}